\documentclass{Interspeech2024}

\usepackage{hyperref}

\usepackage{lipsum}
\usepackage{mwe}
\usepackage{mathtools}
\usepackage{makecell}
\usepackage{cite}
\usepackage{amsmath,amssymb,amsfonts}
\usepackage{pifont}

\usepackage{algorithmic}
\usepackage{graphicx}
\usepackage{textcomp}
\usepackage{xcolor}
\usepackage{wrapfig}
\usepackage{multirow}
\usepackage{tabularx,booktabs}
\usepackage{float}
\usepackage[most]{tcolorbox}
\usepackage{subcaption}
\usepackage{caption}

\usepackage{tikz}
\usetikzlibrary{decorations.pathreplacing,positioning, arrows.meta}
\usetikzlibrary{backgrounds,scopes, chains}

\usepackage{cleveref}
\crefname{section}{Sec.}{Sec.}
\Crefname{section}{Sec.}{Sec.}
\crefname{subsection}{Sec.}{Sec.}
\Crefname{subsection}{Sec.}{Sec.}
\crefname{figure}{Fig.}{Fig.}
\Crefname{figure}{Fig.}{Fig.}
\crefname{equation}{}{}
\Crefname{equation}{}{}
\crefname{table}{Table}{Tables}
\Crefname{table}{Table}{Tables}
\crefname{appendix}{App.}{App.}
\Crefname{appendix}{App.}{App.}

\usepackage{bm}
\usepackage{bbm}
\usepackage{mathrsfs}
\usepackage{amsmath, amssymb}

\newcommand{\C}[1]{}
\def\etal{\textit{et al.}}

\renewcommand*{\vec}[1]{\bm{#1}}
\renewcommand*{\mat}[1]{\bm{#1}}

\newcommand*{\transpose}{^T}

\renewcommand*{\t}{n} 

\newcommand*{\micnum}{M}
\newcommand*{\ch}{m}

\newcommand{\tMic}{y}

\newcommand{\tSrc}{x}

\newcommand*{\Mask}{\mathcal{M}}

\newcommand{\delay}{\tau}
\newcommand*{\est}[1]{\widehat{#1}}

\newcommand{\Feat}{\mathcal{F}}

\newcommand*{\cl}{c}
\newcommand*{\clnum}{\MakeUppercase{\cl}}
\newcommand*{\chcl}{{\ch, \cl}}

\newcommand*{\Coh}{\mathcal{C}}

\newcommand*{\NMF}{B}
\newcommand*{\Eye}{I}
\newcommand*{\Ones}{1}

\interspeechcameraready

\title{Enhanced Deep Speech Separation in Clustered Ad Hoc Distributed Microphone Environments}

\name[affiliation={1}]{Jihyun}{Kim$^{*}$}
\name[affiliation={2}]{Stijn}{Kindt$^{*}$}
\name[affiliation={2}]{Nilesh}{Madhu}
\name[affiliation={1}]{Hong-Goo}{Kang}

\address{
  $^1$Department of Electrical and Electronic Engineering, Yonsei University, South Korea\\
  $^2$IDLab, Ghent University - imec, Ghent, Belgium}
\email{jihyun93815@dsp.yonsei.ac.kr, stijn.kindt@ugent.be, nilesh.madhu@ugent.be, hgkang@yonsei.ac.kr}


\newcommand\blfootnote[1]{%
  \begingroup
  \renewcommand\thefootnote{}\footnote{#1}%
  \addtocounter{footnote}{-1}%
  \endgroup
}

\begin{document}

\maketitle

\begin{abstract}
    Ad-hoc distributed microphone environments, where microphone locations and numbers are unpredictable, present a challenge to traditional deep learning models, which typically require fixed architectures. To tailor deep learning models to accommodate arbitrary array configurations, the Transform-Average-Concatenate (TAC) layer was previously introduced. In this work, we integrate TAC layers with dual-path transformers for speech separation from two simultaneous talkers in realistic settings. However, the distributed nature makes it hard to fuse information across microphones efficiently. Therefore, we explore the efficacy of blindly clustering microphones around sources of interest prior to enhancement. Experimental results show that this deep cluster-informed approach significantly improves the system's capacity to cope with the inherent variability observed in ad-hoc distributed microphone environments.

\end{abstract}

\blfootnote{This work is supported by the Research Foundation - Flanders (FWO) under grant number G081420N\\ \indent $^*$ Equal contributions, shared first author}

\section{Introduction}

\C{The field of ad hoc distributed microphones is starting to receive more attention due to the great spatial coverage it provides. The microphones are connected via a (wireless) link, forming a (wireless) acoustic sensor network (WASN).}

In acoustic sensor networks (ASNs), multiple microphones can be arbitrarily distributed throughout a given space. This distributed configuration provides extensive spatial coverage in contrast to compact microphone arrays \cite{bertrand2011applications}. ASNs are also becoming more common in daily life, with the growing number of devices equipped with one or multiple microphones, like smartwatches, smartphones, laptops and smart glasses. While speech processing \cite{souden2013location, kim23h_interspeech, kim23i_interspeech}, speaker localization \cite{kindt20212d, mpvad2024} and speaker verification \cite{cai2021embedding} have made advancements utilizing this extra spatial information,  there remains much to be explored in this field to {\em fully} capitalize on the possibilities it opens up.

ASNs, particularly those deployed in an ad-hoc manner, have extra challenges associated with them. Firstly, the number of microphones and their respective positions are not known and may vary throughout its operation due to environmental changes. These changes can include devices entering or leaving the environment, or moving within the space. Secondly, due to the potentially widely-distributed nature of the microphones, the same speech signal can be captured at two different microphones at very different time instances. Also, the microphones operated on independent clocks, leading to discrepancies in sample rate offsets (SROs) and sample time offsets (STOs). Lastly, all the microphones could have very different characteristics, i.e. frequency response and directivity. While the latter two can be robustly solved by other methods \cite{gburrek22_sro, chinaev2023long}, the core challenge of speaker separation endures.

One solution previously proposed by Gergen \etal \cite{gergen2018hard,gergen2018fuzzy} is to first cluster the microphones either around the speakers or into a background (noise) cluster. 
Subsequently, cluster information is leveraged within classical signal enhancement frameworks, demonstrating superiority over methods like optimal microphone selection. 
The usefulness of clustering has also been shown in \cite{himawan2010clustered}, where incorporating microphones that are far away from a target speaker (from outside the cluster) can degrade the result. To date, cluster-based separation techniques have only been investigated within the realm of classical signal processing. However, we hypothesize that deep neural network-based separation methods could also benefit from the clustering, potentially surpassing the performance of classical methods.

The Transform-Average-Concatenate (TAC) layer\cite{luo2020end} was previously proposed for deep, array-agnostic, {\em compact} array processing. This layer is introduced between blocks that individually process the inputs for each microphone channel and, consequently, acts as the information sharing layer between microphone channels in a permutation and number invariant manner. Thus, it can handle the challenge of unknown array geometries. TAC has been successfully combined different architectures, e.g., dual-path recurrent neural network (DPRNN)~\cite{luo2020dual}  and VarArray~\cite{yoshioka2022vararray}, where Conformers \cite{gulati2020conformer} are used for time-frequency processing.
Alternative array-agnostic methods, like \cite{wang2020neural}, use multi-head cross-attention to share information across microphones.
However, only limited research \cite{wang2020neural, wang2021continuous} has been conducted specifically on distributed microphone setups. Their investigation revealed both the potential benefits and the intricate challenges associated with employing variably located microphones for speech processing. A significant gap identified is the need for better methods to utilize the spatial diversity of the microphones in a more informed manner.

This paper proposes a novel approach that incorporates the blind microphone clustering techniques into {\em cluster-informed}, array-agnostic deep learning methodologies. In short:
\begin{enumerate}
    \item Initially, in the ad-hoc distributed microphone environment, a blind spatial-statistics-based clustering approach \cite{munoz2021coherence} is employed to cluster microphones around the active speakers. Additionally, the clustering also estimates a {\em pseudo} reference microphone, where the target speech should be the most dominant in all microphones of that cluster.
    \item Then a deep learning-based network exploits spatial information from all microphones within each speaker-dominated cluster to extract the underlying target speech. 
    As we demonstrate, the optimal configuration exploits, in addition to spatial information from all microphones within the cluster, the benefit offered by selecting a robust reference microphone.
\end{enumerate}

Additionally, we propose a training data generation method to simulate clustered data without actually executing the clustering. This is important to save considerable training time and ensures that the training is independent of the specific clustering algorithm employed. The deep separation method will be compared to the classical processing methods and ablation studies will show the effectiveness of the proposed method compared to alternative deep learning structures that do not use the cluster information to its fullest potential. The paper will first overview the classical techniques, where a brief explanation of the clustering and separation method is given in \cref{sec:classical}. Then the proposed deep architecture is explained in \cref{sec:prop_meth}, and evaluated in \cref{sec:experiment}. \cref{sec:conclusion} concludes the paper.

\section{Classical Methods}
\label{sec:classical}
\subsection{Clustering}
\label{sec:clustering}
\begin{figure}[!t] 
    \centering
    {\tikzset{%
	block/.style={line width=0.3mm,draw,align=center,%
		minimum height=4ex%
	},%
	output/.style={block,fill=yellow!40},%
    feat/.style={block,fill=blue!20},%
    clust/.style={block,fill=black!40!green!40},%
    tau/.style={block,fill={rgb:orange,1;yellow,2;pink,5}},%
    dnn/.style={block,fill={rgb:red,3;yellow,2;pink,1}},%
	connect/.style={draw, line width=0.3mm},%
	to/.style={connect,->, line width=0.3mm},%
	from/.style={line width=0.3mm,connect,<-}%
}%

\definecolor{ColorDNN}{RGB}{0, 90, 90}
\definecolor{TextColorDNN}{RGB}{0, 20, 20}

\tikzset{
  shadowtext/.style={
    execute at end picture={
      \fill [color=red, opacity=0] ([xshift=-1ex,yshift=-1ex]current bounding box.south west) rectangle ([xshift=1ex,yshift=1ex]current bounding box.north east);
    }
  }
}

\newcommand{\microphone}{%
    \tikz{
        \begin{scope}
            \clip (-.3em,-.4ex) rectangle (1.3em,1.5ex);
            \fill[black, rounded corners=1.5ex] (-.3em,-.4ex) rectangle (1.3em,5.5ex);
        \end{scope}
        \fill[white, rounded corners=1.3ex] (-.1em,0) rectangle (1.1em,5.1ex);
        \fill[black, rounded corners=1.1ex] (0,.2ex) rectangle (1em,5ex);
        \foreach \pos in {2.5ex, 2.9ex, 3.3ex, 3.7ex}
            \fill[white, rounded corners=.1ex] (.35em,\pos) rectangle +(.8em,.25ex);
        \fill[black] (.4em,-.4ex) rectangle (.6em,-1.5ex);
        \fill[black] (0,-1.5ex) rectangle (1em,-2ex);
    }%
}

\begin{tikzpicture}[node distance=1.4em]%

\fill[color = ColorDNN, opacity = 0.25] (-0.5\columnwidth,-12.25ex) rectangle (0.5\columnwidth, -46.5ex); 
\draw[color=ColorDNN, dashed, line width=0.3mm] (-0.5\columnwidth,-12.25ex) rectangle (0.5\columnwidth, -46.5ex); 
\node[color = TextColorDNN, scale =1.11] at (0, -44ex) {\textbf{Replace with DNN-based array-agnostic processing}};

\node(feat_extr) [feat,minimum width=0.8\columnwidth] {Feature Extraction};

\node (mics) [above=3ex of feat_extr,minimum height=3ex,minimum width=0.3\columnwidth] {\tiny \microphone \hskip5ex \microphone \hskip5ex \huge $\cdots$  \hskip1ex \tiny \microphone};
\draw[to] (mics) -- (feat_extr) node[midway,right,xshift=1ex] {$\micnum \times \tMic_\ch(\t)$};

\node (FCM) [below=3ex of feat_extr,clust,minimum width=0.8\columnwidth] {Clustering Algorithm};
\draw[to] (feat_extr) -- (FCM) node[midway,right,xshift=1ex] {$\micnum \times \Feat$};

\node (mask) [below=6ex of FCM,output,xshift=0.0\columnwidth,minimum width=0.6\columnwidth] {Initial Masking: $\Mask_\cl$};
\node (FCM_mask) [below=3ex of feat_extr,minimum height=3ex,xshift=0.0\columnwidth,minimum width=0.6\columnwidth] {};
\draw[to] (FCM) -- (mask) node[midway,right,xshift=1ex, yshift = 1.5ex] {$\clnum \times$ ref mic};

\node (delay) [below=3ex of mask,tau,minimum width=0.6\columnwidth] {Relative Delay Estimation: $\est{\delay}_\chcl$};
\draw[to] (mask) -- (delay) node[midway,right,xshift=1ex] {$\clnum \times \Mask_\cl$};

\node (DSB) [below=3ex of delay,output,xshift=-0.2125\columnwidth,minimum width=0.375\columnwidth] {DSB};
\node (FCM_DSB) [below=4ex of feat_extr,minimum height=3ex,xshift=-0.38\columnwidth,minimum width=0.2\columnwidth] {};
\node (DSB_FCM) [below=3ex of delay,minimum height=3ex,xshift=-0.38\columnwidth,minimum width=0.2\columnwidth] {};
\draw[to] (FCM_DSB) -- (DSB_FCM) node[midway,right,xshift=1.2ex,yshift=-6.5ex, rotate=90] {hard clusters};
\node (delay_DSB) [below=4ex of mask,minimum height=3ex,minimum width=0.1\columnwidth,xshift=-0.09\columnwidth] {};
\node (DSB_delay) [below=3ex of delay,minimum height=3ex,minimum width=0.6\columnwidth,xshift=-0.09\columnwidth] {};
\draw[to] (delay_DSB) -- (DSB_delay) node[midway,right,xshift=1ex] {$\clnum \times \est{\delay}_\chcl$};

\node (FMVA_DSB) [below=3ex of delay,output,xshift=0.2125\columnwidth,minimum width=0.375\columnwidth] {FMVA-DSB};
\node (FCM_FMVA_DSB) [below=4ex of feat_extr,minimum height=3ex,xshift=0.38\columnwidth,minimum width=0.2\columnwidth] {};
\node (FMVA_DSB_FCM) [below=3ex of delay,minimum height=3ex,xshift=0.38\columnwidth,minimum width=0.2\columnwidth] {};
\draw[to, color = white] (FCM_FMVA_DSB) -- (FMVA_DSB_FCM) node[midway,right,fill=white, scale=0.7, xshift=-1.9ex,yshift=-7.75ex, rotate=90, color=white] {soft clusters};
\draw[to, color = white] (FCM_FMVA_DSB) -- (FMVA_DSB_FCM) node[midway,right,fill=ColorDNN, scale=0.7, xshift=-1.9ex,yshift=-7.75ex, rotate=90, color=ColorDNN, opacity = 0.25] {soft clusters};
\draw[to, opacity = 1] (FCM_FMVA_DSB) -- (FMVA_DSB_FCM) node[midway,right,xshift=-1.5ex,yshift=-6.5ex, rotate=90] {soft clusters};
\node (delay_FMVA_DSB) [below=4ex of mask,minimum height=3ex,minimum width=0.1\columnwidth,xshift=0.09\columnwidth] {};
\node (FMVA_DSB_delay) [below=3ex of delay,minimum height=3ex,minimum width=0.1\columnwidth,xshift=0.09\columnwidth] {};
\draw[to] (delay_FMVA_DSB) -- (FMVA_DSB_delay) node[midway,left,xshift=-2ex,yshift=-4ex] { };

\node (postfilt) [below=3ex of DSB,output,xshift=-0.0\columnwidth,minimum width=0.375\columnwidth] {Postfilter: $\Mask^{\text{Post}}_\cl$};
\node (DSB_postfilt) [below=3ex of DSB,minimum height=3ex,minimum width=0.1\columnwidth] {};
\draw[to] (DSB) -- (DSB_postfilt) node[midway,right,xshift=1ex] {$\clnum \times \est{\tSrc}_{\cl}$};

\node (FMVA_postfilt) [below=3ex of FMVA_DSB,output,xshift=-0.0\columnwidth,minimum width=0.375\columnwidth, dashed] {Postfilter: $\Mask^{\text{Post}}_\cl$};
\node (FMVA_DSB_postfilt) [below=3ex of FMVA_postfilt,minimum height=3ex,minimum width=0.1\columnwidth] {};
\draw[to] (FMVA_DSB) -- (FMVA_postfilt) node[midway,right,xshift=1ex] {$\clnum \times \est{\tSrc}_{\cl}$};


\begin{scope}[on background layer]
    \draw[to, color=gray] (mics.east) -- ++(0.25\columnwidth,0) -- ++(0,-25.25ex) -- (mask.east);
    \draw[to, color=gray] (mics.east) -- ++(0.25\columnwidth,0) -- ++(0,-32.5ex) -- (delay.east);
    \draw[to] (mics.west) -- ++(-0.25\columnwidth,0) -- ++(0,-39.75ex) -- (DSB.west);
    \draw[to] (mics.east) -- ++(0.25\columnwidth,0) -- ++(0,-39.75ex) -- (FMVA_DSB.east);
\end{scope}

\end{tikzpicture}%
}
    \caption{Scheme of the classical cluster-based source separation method. The grayed region is replaced by our proposed separation.\C{The first two blocks are part of the clustering. Then, Inter- and intra-cluster information is exploited to extract the sources dominant in each speech cluster. Yellow blocks indicate stages at which speaker separation can be performed -- and which we use for evaluation. These consist of initial masking, delay and sum beamforming (DSB), fuzzy membership value aware DSB (FMVA-DSB) and postfiltering one of the DSB outputs. The dotted box is a condition that is not included in the tabulated results.}}
    \label{fig:scheme}
    \vspace{-0.25cm}
\end{figure}
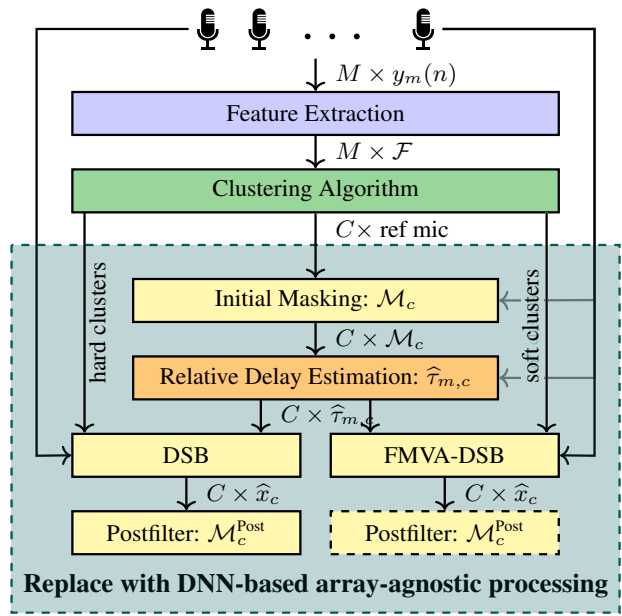

Ensuring robust clustering is essential for distributed microphone techniques~\cite{gergen2015classification,kindt2023robustness}. Based on the findings of the comparative study presented in \cite{kindt2023ad}, the full bandwidth, coherence-based clustering method~\cite{munoz2021coherence} is chosen. This algorithm, represented by the first two steps in \cref{fig:scheme}, uses the pairwise magnitude squared coherence between all $\micnum$ microphones as features, denoted by $\Feat$, and organizes them into a matrix $\mat{\Coh} \in \mathbb{R}^{\micnum \times \micnum}$. Then, non-negative matrix factorization (NMF) \cite{lee2000algorithms} is utilized to cluster the microphones by decomposing this matrix as: $\mat{\Coh} = \mat{\NMF} \mat{\NMF}\transpose \odot (\mat{\Ones} - \mat{\Eye}) + \mat{\Eye}$, where $\odot$ is the element-wise (Hadamard) product, $\mat{\Eye}$ denotes the identity matrix, $\mat{\Ones}$ is the all-ones matrix and $\mat{\NMF} \in \mathbb{R}^{\micnum \times \clnum}$ is the cluster matrix. This matrix 
contains all fuzzy membership values (FMVs) of each microphone towards each cluster, where $\mat{\NMF}_{\ch\cl}$ represents the contribution of microphone $\ch$ to cluster $\cl$. For hard clustering, microphones are then attributed to the cluster where their contribution is highest. Additionally, for each cluster, a reference microphone is identified as the microphone with the highest fuzzy membership value for that cluster. The number of clusters $\clnum$ is one greater than the number of speakers, where the last cluster collects microphones mostly dominated by noise and reverberations. Both the (hard) cluster and the reference microphone will prove invaluable for {\em informed} speech separation. 
\vspace*{-0.2cm}

\subsection{Separation}
\label{ssec.separation}
In \cref{fig:scheme}, the relation between the different classical cluster based separation methods are shown: (1) initial masks are estimated by comparing the amplitude of the short-time Fourier transform (STFT) bins across all reference microphones within each cluster, exploiting sparsity and disjointness of speech \cite{rickard2002approximate}. (2) Then, relative time delays $\est{\delay}_\chcl$ are estimated on the masked cluster signals, and used for delay and sum beamforming (DSB) of all clustered microphone signals. (3) A microphone weighting based on the fuzzy values can be included in the fuzzy membership value aware DSB (FMVA-DSB). (4) Lastly, a postfilter can obtained by comparing the beamformed signal of each cluster. For a detailed overview, we refer to~\cite{gergen2018hard,gergen2018fuzzy}. All the methods succeed in a better foregrounding of target speakers, but the resulting audio quality is poor. The masks of initial and postfilter are binary, distorting the signal, and the simple beamformers cannot suffciently cancel the interferer. 

\begin{figure*}[!t]
\centering
\includegraphics[width=\textwidth]{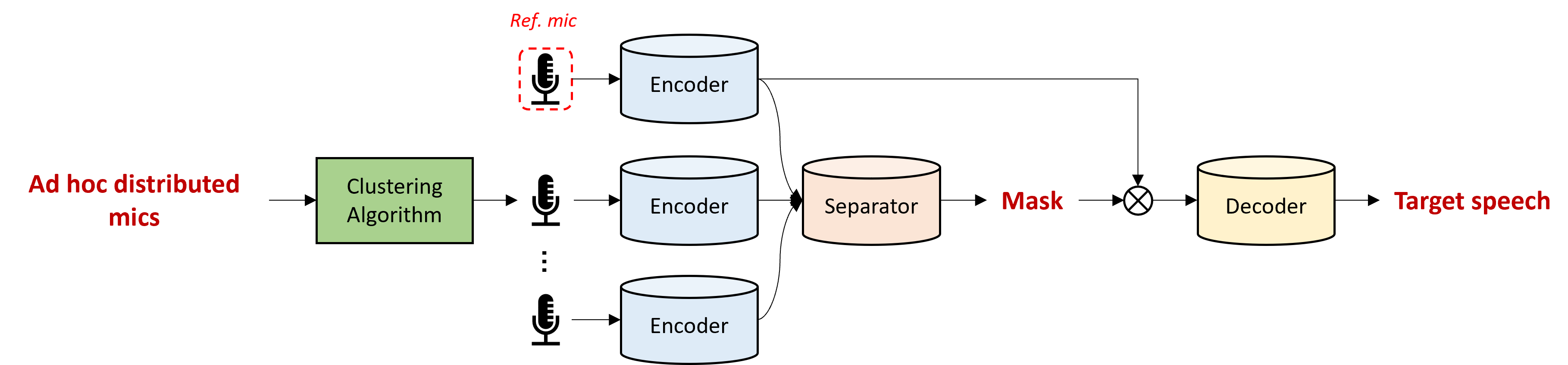}
\caption{Overall system architecture of the proposed deep, array-agnostic, target extraction approach. The reference microphone is identified from the clustering. For separation, the mask is applied to the embedding of the reference microphone.}
\vspace{-15pt}
\label{fig:overall}
\end{figure*}

\section{Proposed Method}
\label{sec:prop_meth}


As illustrated in \cref{fig:overall}, our proposed method clusters ad-hoc distributed microphones around speakers and selects a reference microphone for each cluster. This is done as described in~\cref{sec:clustering}. Microphones of each cluster are then processed 
through a deep learning-based separation network, which consists of Encoder, Separator and Decoder. 

By clustering first, each cluster can be independently processed - removing the need for separation techniques such as permutation invariant training (PIT)~\cite{yu2017permutation}.
We only need to extract the {\em dominant} source of each cluster with a multi-channel time-domain network. The network architecture is based on the VarArray structure \cite{yoshioka2022vararray}, where Conformers are swapped with dual-path transformer networks (DPTNets)~\cite{chen20l_interspeech}. The use of the DPTNet allows for a computationally efficient method to process both local and global information, leading to a comprehensive and accurate representation of the acoustic scene.



\vspace{3pt}
\noindent\textbf{Encoder.}
The Encoder transforms raw multi-channel speech signals $x\in\mathbb{R}^{M\times 1\times T}$ into a high-dimensional feature $h\in\mathbb{R}^{M\times N\times T}$ using a 1-D convolution layer.
\C{\begin{align}
    \label{eq:encoder}
    h &= \mathrm{ReLU}(conv1d(x)).
\end{align}}
Additionally, the Encoder incorporates a segmentation strategy derived from DPTNet, allowing for more precise handling of both local and global dependencies within the speech signal.
It splits the hidden feature $h$ into overlapped chunks of length $K$ with a hop size of $K/2$. The hidden feature $h$ is thus a 4-D tensor $h_{0}\in\mathbb{R}^{M\times N\times K\times P}$, where $P$ is the number of chunks.
This design choice ensures that the network both preserves the detailed temporal structure of the speech signal and increases its receptive field. This is essential to compensate for the relatively long time delays in widely distributed microphone settings.


\noindent\textbf{Separator.}
The separator combines TAC layers and DPTNets within its processing chain for spatial and temporal processing respectively. Similar to the structure of VarArray, 3 DPTNets are interleaved with 2 TAC layers, followed by a mean pooling and 2 DPTNets. The first DPTNets process each microphone individually, where the TAC layers combine the microphone information. Mean pooling reduces computational complexity for the following (single channel) DPTNets. The separator produces a mask, which is applied to an encoder embedding. Here the suitability of the clustering, more specifically the indication of a reference microphone, is again clear. If this information is present, the hidden feature of the reference microphone can be selected for masking. 

\begin{figure}[!t]
\centering
\includegraphics[width=0.5\textwidth]{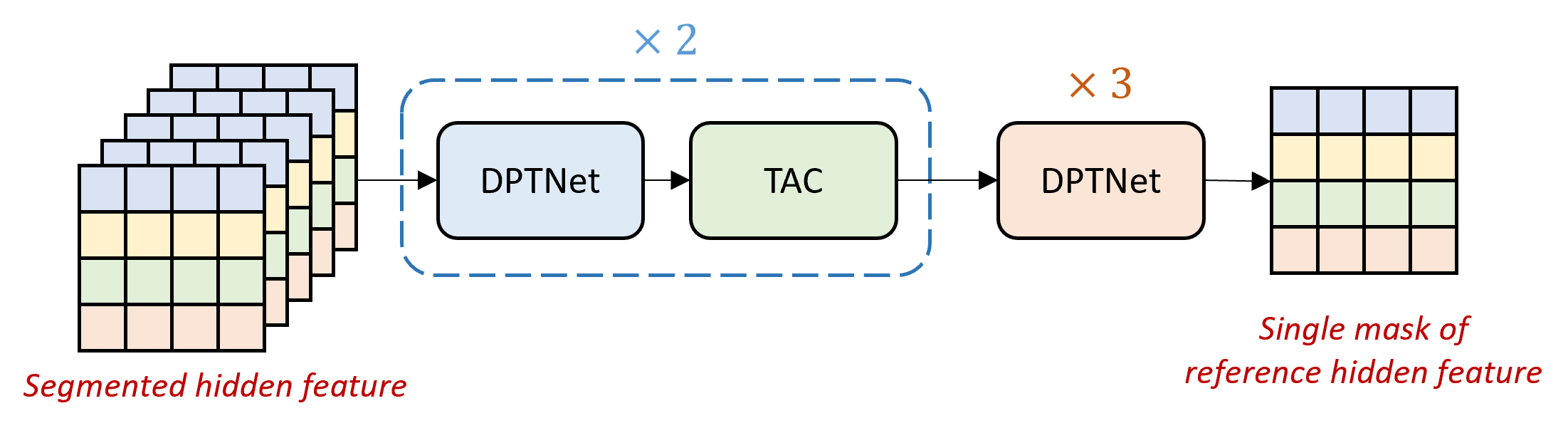}
\caption{Detail structure of the separator. 
For multi-microphone processing, a single module operates in parallel across the microphones. }
\label{fig:separator}
\vspace{-15pt}
\end{figure}

We propose further changing the mean pooling layer by selecting the embedding from the reference microphone. Firstly, this embedding should be the most dominated by the target speech. Secondly, it reduces the computational cost since no DPTNets operations are applied on the non reference embeddings. The proposed method is depicted in \cref{fig:separator}. We will term this proposed style, while the previously described networks VarArray style in the ablation studies of \cref{sec:results}.



\vspace{4pt}
\noindent\textbf{Decoder.}
The Decoder reconstructs the separated speech signals from the enhanced high-dimensional latent representation.
In a process that mirrors the encoder, the decoder employs overlap and add, followed by a 1-D transposed convolution layer, to transform the enhanced hidden feature back into the time-domain yielding clear, distinct speech tracks.


\section{Experimental Evaluation}
\label{sec:experiment}

\begin{figure*}[tbh]
    \vskip-6ex
    \centering
    \begin{subfigure}{0.32\textwidth}
    \centering
        \includegraphics[width=\textwidth, trim=0 50 10 40, clip]{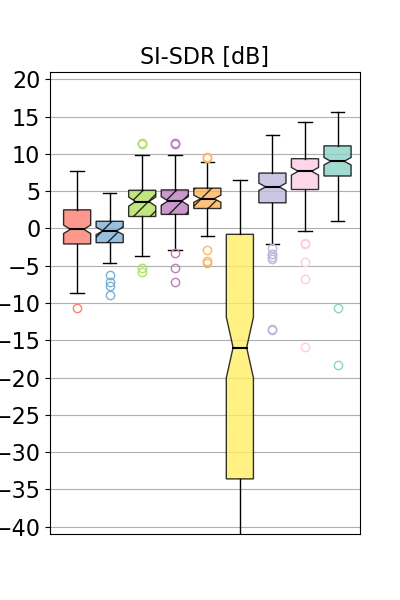}
    \end{subfigure}
    \begin{subfigure}{0.32\textwidth}
    \centering
        \includegraphics[width=\textwidth, trim=0 50 10 40, clip]{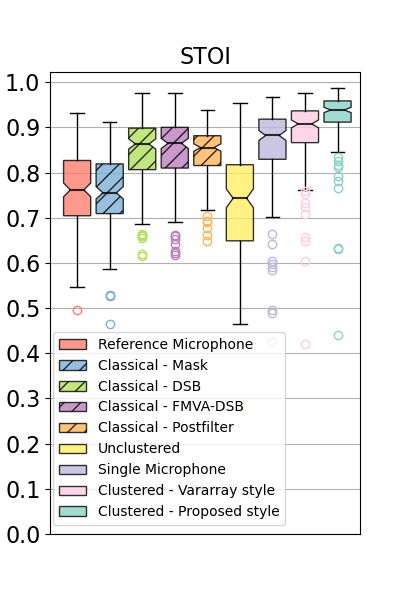}
    \end{subfigure}
    \begin{subfigure}{0.32\textwidth}
    \centering
        \includegraphics[width=\textwidth, trim=0 50 10 40, clip]{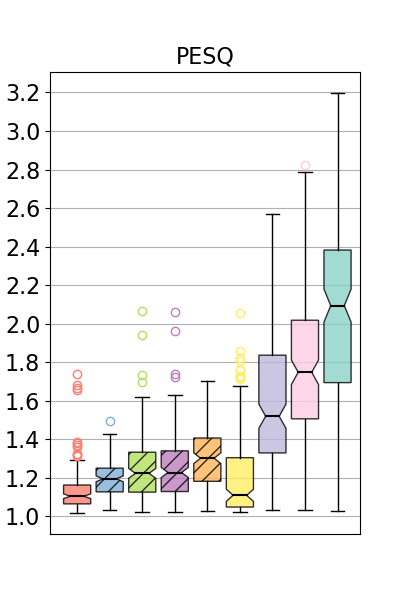}
    \end{subfigure}

    \vskip-1ex
    \caption{SI-SDR, STOI and PESQ for the different separation methods (higher is better)}
    \vskip-4ex
\label{fig:separation_metrics}
\end{figure*}

In this section, we want to show that deep networks improve upon classical separation methods and that cluster information is essential for good separation. Therefore, next to comparison with the classical methods, we include ablation studies where all microphones are used as input to the neural network ({\em unclustered version}), an ablation study where only the reference microphone ({\em single microphone}) is used as input, and study the difference between VarArray style and the proposed style.

The unclustered version is trained with PIT loss to separate the different speakers. Since no reference mic is known, a random microphone is selected of which the encoding embeddings are taken for the decoder. Initial experiments showed that averaging over the embeddings performs worse. This highlights why clustering in widely distributed microphones is beneficial. 

The ablation study with the single microphone method, where no TAC layers are needed, indicates whether the spatial diversity, provided by the multiple microphones within a cluster, is valuable. The VarArray style ablation study reveals whether incorporating the reference microphone information within the network structure improves the final result.

We will begin by outlining the datasets used for training and evaluation, as well as the specific parameter choices, before delving into the results.
\vskip-2ex

\subsection{Dataset}
\vskip-1ex

To train our network, we utilized the WSJ0-2mix \cite{hershey2016deep} clean speech dataset, and convolved them with the shoebox room impulse responses (RIRs), generated by the image source model of gpuRIR \cite{diaz2021gpurir}. White noise was added at SNRs uniformly sampled between 0 dB and 20 dB. A variety of different rooms (dimensions and reverberation times) are generated to promote generalizability, totaling 10,080 different scenarios. For this work, the simulation was limited to cases where the two speakers were located in different halves of the room. A total of 16 microphones were simulated for each scenario, where for each source there are at least 3 microphones within its critical distance, consistent with previous cluster based separation work \cite{kindt2023robustness}. We iterate over all RIRs for each epoch and select a random clean speech sample on the fly to increase diversity. Also, for each batch, a random selection of microphones -- between 8 and 16 microphones -- is chosen, to expose the network to different numbers of microphones and increase the total number of possible scenarios. However, it is ensured that the microphones within the critical distance are kept during selection.

\C{
a widely recognized benchmark for speech separation tasks. The speakers are put in a room and captured by a distributed microphone by convolving the signal with room impulse responses (RIRs), generated by gpuRIR \cite{diaz2021gpurir}.
To tailor our network for adaptability to various room configurations and microphone placements, we generated RIRs from 10 distinct room settings with different room dimensions and reverberation times.
In each room, 144 locations for speaker pairs were randomly selected, for which 7 distinct distributed microphone configurations were sampled. During this location sampling, it was made sure that one speaker was positioned in the left half of the room, while the other was in the right half of the room, consistent with prior work \cite{}. This is done to ensure that the speakers are not too close to each other. Additionally, for each speaker, there are at least 3 microphones put in the critical distance of the speaker and 10 microphone positions are randomly sampled, which is also consistent with previous clustering work \cite{}. This totals 10,080 distinct scenarios where, during training, a random number of microphones between 3 and 7 will be picked. This lets the network experience various numbers of input channels during training.
}
\noindent\textbf{Clustered Training Dataset.}
However, clustering on the fly would waste valuable training time since NMF is an iterative method. Also, this might make the network dependent on the clustering algorithm. To alleviate these problems, a second dataset of clustered RIRs is generated, where microphone positions are simulated as if they could have originated from clustering. Two sources are still sampled in different room halves, but the microphones are no longer sampled in the whole room. Three microphones are still placed within the critical distance of the speaker, and one is selected as the reference microphone. 4 other microphones are simulated in a $2m \times 2m$ square centered around the speaker. During training, a random number of microphones between 3 and 7 is chosen for generalizability to unknown microphone numbers. The unclustered dataset is utilized to train the unclustered version, while the clustered dataset is used to train all other networks.



\noindent\textbf{Evaluation Dataset.}
To assess the real-world applicability of the model, the realistic SINS dataset \cite{SINS2023Glitza}, simulated with a CATT model, is used. The evaluation set is done similarly to~\cite{kindt2023robustness, kindt2023ad}.Two speaker positions are selected in opposite halves of the room and 16 microphones are distributed over the room. For each source, at least 3 microphones are within the critical distance. If the speakers are both sampled towards the middle of the room, they {\em could} still be very closely spaced. Dry speech is taken from the LibriSpeech dataset \cite{panayotov2015librispeech}. White noise is added, at an SNR of 10 dB with respect to the middle of the room. The actual SNR at individual microphones can differ greatly.

Good performance on real clustered data would validate the clustered training dataset. Also, since this dataset differs significantly in realism and sound sources, it can demonstrate the model's performance and generalization capability for environments that closely mimic actual speech separation challenges.

\vspace{-5pt}
\subsection{Experiment Setup}
For the Encoder and Decoder, we selected a kernel size of 8 samples with a stride of 50$\%$. We use a segment size $K$ of 250 on the segmentation for dual-path processing. We set the feature dimension of the separation network to be 64. We use 4 attention heads on each transformer layer in the DPTNet.
The training criterion we used is the Scale Invariant Signal to Distortion Ratio (SI-SDR) \cite{le2019sdr} loss \C{, aiming to maximize the SI-SDR by minimizing the loss function which is defined as:
\begin{align}
    \label{eq:criterion}
    \mathcal{L}_{SI-SDR}
    &= -10 \log_{10} \dfrac{|| \alpha s||^{2}}{|| \alpha s - \Hat{s}||^{2}}. 
\end{align}
where $s, \Hat{s}$ are, respectively, the clean and estimated speech signal and $\alpha$ is a scaling factor that aligns the estimate to the target signal, calculated as $\alpha = (\Hat{s}^{T} \! \cdot s)/ ||s||^{2}$.} We used Adam optimizer and the training process began with an initial learning rate of 0.125, with a strategy to halve the learning rate if the validation loss doesn't decrease for three epochs. 
The total number of parameters in our proposed model is 2.23 M.

\vspace{-5pt}
\subsection{Experiment Results}
\label{sec:results}

To assess the performance, we employed three objective metrics: Scale Invariant Signal to Distortion Ratio (SI-SDR) \cite{le2019sdr}, Perceptual Evaluation of Speech Quality (PESQ) \cite{rix2001perceptual}, and Short-Time Objective Intelligibility (STOI) \cite{taal2010short}

\cref{fig:separation_metrics} shows the results of the different separation methods. The {\em reference point} is given by the metrics computed on the {\em unprocessed} reference microphone signals. Firstly, it is clear, from the SI-SDR and PESQ, that combining information across {\em all} microphones (unclustered version) does not perform well. This indicates that the general, uninformed nature of TAC -- which combines features from all microphone signals similarly, independent of the underlying speaker or background noise dominance at that microphone -- does not suit distributed scenarios. Additionally, the lack of a good selection mechanism for hidden features on which the mask is applied, makes it hard for the network to generalize to other situations: the SI-SDR performance is very poor even though it was trained to maximize this metric.

Picking the unprocessed microphone closest to each speaker -- the reference microphone in the clustering algorithm --  outperforms the unclustered deep leaning method and shows decent intelligibility (STOI). However, the output can be further improved. The classical methods do indeed increase the performance on all metrics, except for the initial masking -- which is anyway mainly used for a robust, relative time delay estimation.

Using cluster informed deep learning algorithms significantly outperforms the classical methods. Using the reference microphone as input for a single channel model, where only DPTNets are sequentially applied, gives a big performance boost, most notably the big increase in PESQ. However, the method does not exploit the spatial diversity provided by the clustered microphones. When considering the inputs from clustered microphones, the metrics show that, unlike the unclustered version, TAC is effective. All microphones are dominated by the same source, removing target ambiguity. The performance of the clustered methods also supports the validity of the proposed data generation scheme.

The results also show that it is worthwhile to let the network prioritize the embeddings from the reference microphone. VarArray style averages the features over all the microphones before continuing with the single channel portion of the network. The proposed style takes the reference microphone as input for the single-channel portion of the network. This simple information inclusion in the design increases the performance significantly on all three metrics.

Specific scenarios, where the clusters are plotted and audio of the different separation techniques is present, can be found at https://aspire.ugent.be/demos/INTERSPEECH2024SK/ .
\vspace{-5pt}
\section{Conclusion}
\label{sec:conclusion}
\vspace{-5pt}

In this paper, we introduced a novel approach for speech separation in ad hoc distributed microphone environments, combining coherence-based clustering methods with deep learning networks.
Our experiments on realistically simulated RIRs show that it is essential to include cluster information in deep learning separation networks. More so, also including the reference microphone -- a byproduct of the clustering method -- further enhances the method. Conversely, the deep learning based separation gives a significant boost to the separation compared to classical methods. This highlights the benefits of combining traditional signal processing techniques with modern deep learning for speech processing tasks in real-world scenarios. Additionally, an efficient data generation paradigm to simulate clustered data was proposed for training such frameworks. 

Future work could further increase the information the networks get from the clustering, by incorporating cross cluster information within the design of the network.


{
\small
\bibliographystyle{MyIEEEtran}
\bibliography{references}

\begin{thebibliography}{10}
\providecommand{\url}[1]{#1}
\csname url@samestyle\endcsname
\providecommand{\newblock}{\relax}
\providecommand{\bibinfo}[2]{#2}
\providecommand{\BIBentrySTDinterwordspacing}{\spaceskip=0pt\relax}
\providecommand{\BIBentryALTinterwordstretchfactor}{4}
\providecommand{\BIBentryALTinterwordspacing}{\spaceskip=\fontdimen2\font plus
\BIBentryALTinterwordstretchfactor\fontdimen3\font minus \fontdimen4\font\relax}
\providecommand{\BIBforeignlanguage}[2]{{%
\expandafter\ifx\csname l@#1\endcsname\relax
\typeout{** WARNING: IEEEtran.bst: No hyphenation pattern has been}%
\typeout{** loaded for the language `#1'. Using the pattern for}%
\typeout{** the default language instead.}%
\else
\language=\csname l@#1\endcsname
\fi
#2}}
\providecommand{\BIBdecl}{\relax}
\BIBdecl

\bibitem{bertrand2011applications}
A.~Bertrand, ``Applications and trends in wireless acoustic sensor networks: A signal processing perspective,'' in \emph{2011 18th IEEE symposium on communications and vehicular technology in the Benelux (SCVT)}.\hskip 1em plus 0.5em minus 0.4em\relax IEEE, 2011, pp. 1--6.

\bibitem{souden2013location}
M.~Souden, K.~Kinoshita, M.~Delcroix, and T.~Nakatani, ``Location feature integration for clustering-based speech separation in distributed microphone arrays,'' \emph{IEEE/ACM Transactions on Audio, Speech, and Language Processing}, vol.~22, no.~2, pp. 354--367, 2013.

\bibitem{kim23h_interspeech}
D.~Kim, S.-W. Chung, H.~Han, Y.~Ji, and H.-G. Kang, ``{HD-DEMUCS: General Speech Restoration with Heterogeneous Decoders},'' in \emph{Proc. INTERSPEECH 2023}, 2023, pp. 3829--3833.

\bibitem{kim23i_interspeech}
J.~Kim and H.-G. Kang, ``{Contrastive Learning based Deep Latent Masking for Music Source Separation},'' in \emph{Proc. INTERSPEECH 2023}, 2023, pp. 3709--3713.

\bibitem{kindt20212d}
S.~Kindt, A.~Bohlender, and N.~Madhu, ``2d acoustic source localisation using decentralised deep neural networks on distributed microphone arrays,'' in \emph{Speech Communication; 14th ITG Conference}.\hskip 1em plus 0.5em minus 0.4em\relax VDE, 2021, pp. 1--5.

\bibitem{mpvad2024}
H.~Han and N.~Kumar, ``A cross-talk robust multichannel vad model for multiparty agent interactions trained using synthetic re-recordings,'' in \emph{2024 Hands-free Speech Communications and Microphone Arrays (HSCMA)}, 2024.

\bibitem{cai2021embedding}
D.~Cai and M.~Li, ``Embedding aggregation for far-field speaker verification with distributed microphone arrays,'' in \emph{2021 IEEE spoken language technology workshop (SLT)}.\hskip 1em plus 0.5em minus 0.4em\relax IEEE, 2021, pp. 308--315.

\bibitem{gburrek22_sro}
T.~Gburrek, J.~Schmalenstroeer, and R.~Haeb-Umbach, ``On synchronization of wireless acoustic sensor networks in the presence of time-varying sampling rate offsets and speaker changes,'' in \emph{IEEE International Conference on Acoustics, Speech and Signal Processing (ICASSP)}, 2022, pp. 916--920.

\bibitem{chinaev2023long}
A.~Chinaev, N.~Knaepper, and G.~Enzner, ``Long-term synchronization of wireless acoustic sensor networks with nonpersistent acoustic activity using coherence state,'' in \emph{ICASSP 2023-2023 IEEE International Conference on Acoustics, Speech and Signal Processing (ICASSP)}.\hskip 1em plus 0.5em minus 0.4em\relax IEEE, 2023, pp. 1--5.

\bibitem{gergen2018hard}
S.~Gergen, R.~Martin, and N.~Madhu, ``Source separation by feature-based clustering of microphones in ad hoc arrays,'' in \emph{2018 16th International Workshop on Acoustic Signal Enhancement (IWAENC)}.\hskip 1em plus 0.5em minus 0.4em\relax IEEE, 2018, pp. 530--534.

\bibitem{gergen2018fuzzy}
S.~Gergen, R.~Martin, and N.~Madhu, ``Source separation by fuzzy-membership value aware beamforming and masking in ad hoc arrays,'' in \emph{Speech Communication; 13th ITG-Symposium}.\hskip 1em plus 0.5em minus 0.4em\relax VDE, 2018, pp. 1--5.

\bibitem{himawan2010clustered}
I.~Himawan, I.~McCowan, and S.~Sridharan, ``Clustered blind beamforming from ad-hoc microphone arrays,'' \emph{IEEE Transactions on Audio, Speech, and Language Processing}, vol.~19, no.~4, pp. 661--676, 2010.

\bibitem{luo2020end}
Y.~Luo, Z.~Chen, N.~Mesgarani, and T.~Yoshioka, ``End-to-end microphone permutation and number invariant multi-channel speech separation,'' in \emph{ICASSP 2020-2020 IEEE International Conference on Acoustics, Speech and Signal Processing (ICASSP)}.\hskip 1em plus 0.5em minus 0.4em\relax IEEE, 2020, pp. 6394--6398.

\bibitem{luo2020dual}
Y.~Luo, Z.~Chen, and T.~Yoshioka, ``Dual-path rnn: efficient long sequence modeling for time-domain single-channel speech separation,'' in \emph{ICASSP 2020-2020 IEEE International Conference on Acoustics, Speech and Signal Processing (ICASSP)}.\hskip 1em plus 0.5em minus 0.4em\relax IEEE, 2020, pp. 46--50.

\bibitem{yoshioka2022vararray}
T.~Yoshioka, X.~Wang, D.~Wang, M.~Tang, Z.~Zhu, Z.~Chen, and N.~Kanda, ``Vararray: Array-geometry-agnostic continuous speech separation,'' in \emph{ICASSP 2022-2022 IEEE International Conference on Acoustics, Speech and Signal Processing (ICASSP)}.\hskip 1em plus 0.5em minus 0.4em\relax IEEE, 2022, pp. 6027--6031.

\bibitem{gulati2020conformer}
A.~Gulati, J.~Qin, C.-C. Chiu, N.~Parmar, Y.~Zhang, J.~Yu, W.~Han, S.~Wang, Z.~Zhang, Y.~Wu \emph{et~al.}, ``Conformer: Convolution-augmented transformer for speech recognition,'' \emph{Interspeech 2020}, 2020.

\bibitem{wang2020neural}
D.~Wang, Z.~Chen, and T.~Yoshioka, ``Neural speech separation using spatially distributed microphones,'' \emph{INTERSPEECH 2020}, pp. 2467--2471, 2020.

\bibitem{wang2021continuous}
D.~Wang, T.~Yoshioka, Z.~Chen, X.~Wang, T.~Zhou, and Z.~Meng, ``Continuous speech separation with ad hoc microphone arrays,'' in \emph{2021 29th European Signal Processing Conference (EUSIPCO)}.\hskip 1em plus 0.5em minus 0.4em\relax IEEE, 2021, pp. 1100--1104.

\bibitem{munoz2021coherence}
A.~J. Mu{\~n}oz-Montoro, P.~Vera-Candeas, and M.~G. Christensen, ``A coherence-based clustering method for multichannel speech enhancement in wireless acoustic sensor networks,'' in \emph{2021 29th European Signal Processing Conference (EUSIPCO)}.\hskip 1em plus 0.5em minus 0.4em\relax IEEE, 2021, pp. 1130--1134.

\bibitem{gergen2015classification}
S.~Gergen, A.~Nagathil, and R.~Martin, ``Classification of reverberant audio signals using clustered ad hoc distributed microphones,'' \emph{Signal Processing}, vol. 107, pp. 21--32, 2015.

\bibitem{kindt2023robustness}
S.~Kindt, J.~Thienpondt, L.~Becker, and N.~Madhu, ``Robustness of ad hoc microphone clustering using speaker embeddings: evaluation under realistic and challenging scenarios,'' \emph{EURASIP Journal on Audio, Speech, and Music Processing}, vol. 2023, no.~1, p.~46, 2023.

\bibitem{kindt2023ad}
S.~Kindt, M.~Meeldijk, and N.~Madhu, ``Ad hoc distributed microphones clustering: A comparative analysis on using coherence and signal-specific features,'' in \emph{Speech Communication; 15th ITG Conference}.\hskip 1em plus 0.5em minus 0.4em\relax VDE, 2023, pp. 11--15.

\bibitem{lee2000algorithms}
D.~Lee and H.~S. Seung, ``Algorithms for non-negative matrix factorization,'' \emph{Advances in neural information processing systems}, vol.~13, 2000.

\bibitem{rickard2002approximate}
S.~Rickard and O.~Yilmaz, ``On the approximate {W}-disjoint orthogonality of speech,'' in \emph{2002 IEEE International Conference on Acoustics, Speech, and Signal Processing (ICASSP)}, vol.~1.\hskip 1em plus 0.5em minus 0.4em\relax IEEE, 2002, pp. I--529.

\bibitem{yu2017permutation}
D.~Yu, M.~Kolb{\ae}k, Z.-H. Tan, and J.~Jensen, ``Permutation invariant training of deep models for speaker-independent multi-talker speech separation,'' in \emph{2017 IEEE International Conference on Acoustics, Speech and Signal Processing (ICASSP)}.\hskip 1em plus 0.5em minus 0.4em\relax IEEE, 2017, pp. 241--245.

\bibitem{chen20l_interspeech}
J.~Chen, Q.~Mao, and D.~Liu, ``{Dual-Path Transformer Network: Direct Context-Aware Modeling for End-to-End Monaural Speech Separation},'' in \emph{Proc. Interspeech 2020}, 2020, pp. 2642--2646.

\bibitem{hershey2016deep}
J.~R. Hershey, Z.~Chen, J.~Le~Roux, and S.~Watanabe, ``Deep clustering: Discriminative embeddings for segmentation and separation,'' in \emph{2016 IEEE international conference on acoustics, speech and signal processing (ICASSP)}.\hskip 1em plus 0.5em minus 0.4em\relax IEEE, 2016, pp. 31--35.

\bibitem{diaz2021gpurir}
D.~Diaz-Guerra, A.~Miguel, and J.~R. Beltran, ``gpurir: A python library for room impulse response simulation with gpu acceleration,'' \emph{Multimedia Tools and Applications}, vol.~80, pp. 5653--5671, 2021.

\bibitem{SINS2023Glitza}
R.~Glitza, L.~Becker, A.~Nelus, and R.~Martin, ``Database of simulated room impulse responses for acoustic sensor networks deployed in complex multi-source acoustic environments.'' in \emph{EUSIPCO}, 2023.

\bibitem{panayotov2015librispeech}
V.~Panayotov, G.~Chen, D.~Povey, and S.~Khudanpur, ``Librispeech: an asr corpus based on public domain audio books,'' in \emph{2015 IEEE international conference on acoustics, speech and signal processing (ICASSP)}.\hskip 1em plus 0.5em minus 0.4em\relax IEEE, 2015, pp. 5206--5210.

\bibitem{le2019sdr}
J.~Le~Roux, S.~Wisdom, H.~Erdogan, and J.~R. Hershey, ``Sdr--half-baked or well done?'' in \emph{ICASSP 2019-2019 IEEE International Conference on Acoustics, Speech and Signal Processing (ICASSP)}.\hskip 1em plus 0.5em minus 0.4em\relax IEEE, 2019, pp. 626--630.

\bibitem{rix2001perceptual}
A.~W. Rix, J.~G. Beerends, M.~P. Hollier, and A.~P. Hekstra, ``Perceptual evaluation of speech quality ({PESQ})-a new method for speech quality assessment of telephone networks and codecs,'' in \emph{IEEE Intl. Conf. on acoustics, speech, and signal processing.}, vol.~2, 2001, pp. 749--752.

\bibitem{taal2010short}
C.~H. Taal, R.~C. Hendriks, R.~Heusdens, and J.~Jensen, ``A short-time objective intelligibility measure for time-frequency weighted noisy speech,'' in \emph{IEEE Intl. Conf. on acoustics, speech and signal processing}, 2010, pp. 4214--4217.

\end{thebibliography}
}

\end{document}